\documentclass[preprint,12pt,showkeys,nofootinbib,floatfix]{revtex4}
\usepackage{amssymb}
\usepackage{amsfonts}
\usepackage{amsmath}
\usepackage{graphicx}
\usepackage{subfigure}
\usepackage[dvipdfm,
            pdfstartview=FitH,
            bookmarksnumbered=true,
            bookmarksopen=true,
            colorlinks,
            pdfborder=001,
            linkcolor=green,
            anchorcolor=green,
            citecolor=red
            ]{hyperref}
\usepackage[toc,page,title,titletoc,header]{appendix}
\usepackage{epsfig}
\usepackage{slashed}
\usepackage{latexsym}
\usepackage{syntonly}
\newcommand{\be}{\begin{equation}}
\newcommand{\bea}{\begin{eqnarray}}
\newcommand{\eea}{\end{eqnarray}}
\newcommand{\ba}{\begin{array}}
\newcommand{\ea}{\end{array}}
\newcommand{\ee}{\end{equation}}

\def\Bp#1{{\bar{#1}}^{\prime}}  
\def\dint{\mathop{\idotsint}}
\begin{document}
\title{Several integrals of  quaternionic field on hyperbolic matrix space}
\author{Fu-Wen Shu$^{1,2,3}$}
\thanks{E-mail address: shufuwen@ncu.edu.cn}
\author{You-Gen Shen$^{3}$}
\thanks{E-mail address: ygshen@shao.ac.cn}
\affiliation{
$^{1}$Department of Physics, Nanchang University, Nanchang, 330031, China\\
$^{2}$Center for Relativistic Astrophysics and High Energy Physics, Nanchang University, Nanchang, 330031, China\\
$^{3}$ Shanghai Astronomical observatory, Chinese Academy of Sciences, Shanghai, 200030, China}
\begin{abstract}
Some integrals of matrix spaces over a quaternionic field have been calculated in this work. The associated volume  of hyperbolic matrix spaces over a quaternionic field has also been calculated by making use of these integrals, and it is of great significance in calculating related kernel functions of these spaces.
\end{abstract}

\keywords{Quaternionic field, Integral on matrix spaces, Hyperbolic matrix space}
\maketitle
\section{Introduction}
Matrices over a ring, as pointed out by Hua in \cite{hua2}, deserve particular attention due to their universalities and their close relationship to matrix theory over some special domains.

Quaternion plays an important role in modern mathematics, it forms a $4$-dimensional associative division algebra over the real domain. It actually  according to the Frobenius theorem\cite{frobenius,palais}, is one of only two finite-dimensional division rings containing the real numbers as a proper subring (the other is the complex number), and quaternions are the largest Euclidean Hurwitz algebra among these rings. Moreover, the quaternions were the first noncommutative division algebra to be discovered. It is therefore of particular importance in analysis. In addition, quaternion is also useful in applied mathematics. Particular example is calculations involving three-dimensional rotations such as in three-dimensional computer graphics and crystallographic texture analysis.

Harmonic analysis and harmonic functions on classical domains over  complex numbers have been fully studied by Hua in \cite{Hua1,hua3}. Ref. \cite{lu1} also gave brief discussions on real domain for associated classical domains. Refs. \cite{zheng} and \cite{ms} made, respectively, a full investigation on several classical groups (unitary group, orthogonal group and unitary  symplectic group) and harmonic analysis on compact Lie groups and compact homogeneous space.

It was first noted by Hua that $USp[2n]$ actually is nothing but unitary group over a quaternionic field, and it can be viewed as a classical manifold of the first classical domain over the quaternionic field,
\be
\Re_I(Q)=\left\{Q=(q_{ij}):\  I^{m}-Q\Bp{Q}>0, \ Q=Q^{(m,n)}\right\},
\ee
where $q_{ij}$ are quaternions.  Besides $\Re_I$, there are another two quaternionic classical domains which can be defined as following
\bea
\Re_{II}(Q)&=&\left\{Q=(q_{ij}):\  I-Q^2>0, \ Q=Q^{(n)}=\bar Q'\right\},\\
\Re_{III}(Q)&=& \left\{Q=(q_{ij}):\  I+Q^2>0, \ Q=Q^{(n)}=-\bar Q'\right\}.
\eea
 It was shown in ref. \cite{sun} that these three quaternionic classical domains correspond, respectively, to equivalent representations of $Sp(m,n)/Sp(m)\times Sp(n)$, $SU^*(2n)/Sp(m)\times Sp(n)$\footnote{For $\Re_{II}$, one additional condition should be fulfilled, i.e., $\det(I+Q)=\det(I-Q)$.}, $Sp(n,\mathbb{C})/Sp(m)\times Sp(n)$ in E Cartan's list of irreducible Riemannian globally symmetric spaces\cite{helgason}.
 
 Calculations of Bergmann kernel, Cauchy kernel and Poisson kernel in symmetric classical domains are of fundamental importance in theories of analytic functions of several variables\cite{ay,kz1,kz2}. In order to calculate these kernel functions and volumes of the associated matrix spaces, we should obtain some related integrals on these associated matrix spaces\cite{Hua1,hua3}. This is one of the main motivations of this work.
 
 In this paper, we calculate several integrals on matrix spaces over a quaternionic field. With these integrals, we obtain volumes of the associated hyperbolic matrix spaces over a quaternionic field. This will be very useful in the future calculations for kernel functions.
 
 In addition, our another motivation of this work comes from some recent progress in theoretical physics, particularly in string theory, gravitational  physics and cosmology. It is well known that many ideas and techniques underlying in the study of the classical domains can shed lights on the study of the theoretical physics. Actually, in the early 1970s, there was a proposal that generalizes the Einstein's special relativity to the model with nonzero cosmological constant \cite{lu2}. In this work, they also discussed the possible kinematic effects in the classical domains and the red-shift
phenomena in our universe. Recently applications of this field was revived after the discovery of the accelerating expansion of our universe in the late 1990s\cite{perlummutter,perlummutter1,bennet}. Many generalizations of general relativity in the framework of classical domains have been done, see \cite{guo1,guo2,guo3,guo4,guo5,guo6,guo7,guo8,guo9} for an incompete list.
 
 Another application which attracts less enough attention is that the classical domains may play a particular significance in the study of AdS/CFT correspondence, which states that string theory in the AdS space is holographically dual to a CFT on the boundary of the AdS\cite{maldacena}. Witten and Yau  later fully investigated the possible proof of the AdS/CFT conjecture  in \cite{witten,yau}. But their attempts were based on the Euclidean version of the AdS, and focused only on massless scalar field. It was Chang \textit{et al}  who first noted in \cite{cg,lu3}  that the Poisson kernels and relations between them within the framework of the classical manifolds and classical domains can be used to prove the AdS/CFT conjecture even for massive scalar field with the right signature. They generalized their proof to the massive Dirac field in \cite{lu4}. All these progresses show that it is of great importance in the applications of classical manifolds and classical domains to theoretical physics. This is due to the remarkable relations between AdS which is a  classical manifold, and its boundary CFT which is a sub manifold of the associated classical manifold--- the so-called extended space of Lie-sphere named by Hua, which is also a classical manifold. 

\section{Preliminary}
The algebra of quaternions $\Bbb Q$ is a unital $\Bbb R$-algebra
with generators $\bold i$, $\bold j$, $\bold k $ such that
$$
\bold i^2=\bold j^2=\bold k^2=-1, \ \
\bold i \bold j=\bold k = -\bold j \bold i, \ \
\bold j \bold k=\bold i=-\bold k \bold j, \ \
\bold k \bold i=\bold j=-\bold i \bold k.
$$

The quaternions set $\Bbb Q$ can be mapped to to $\Bbb R^4$, a four-dimensional vector space over the real numbers, whose basis customarily denoted as $1,\bold  i,\bold j,$ and $\bold k$. Every element of $\Bbb Q$ can be uniquely written as a linear combination of these basis elements, that is, for $q\in \Bbb Q$ one has $q=\alpha+\beta\bold i + \gamma\bold
j+\delta\bold k $ where $\alpha,\beta,\gamma,\delta\in \Bbb R$. 
Conjugation of quaternions is analogous to conjugation of complex numbers and to transposition of elements of Clifford algebras. That is, for any
$q\in\Bbb Q$, we have $\bar q=\alpha-\beta\bold i - \gamma\bold
j-\delta\bold k  $ and satisfies $\overline {q_1q_2}=\bar q_2\bar q_1$. It
follows that $q\bar q=\bar q q$.

With the definition of the conjugation, one can define norm $\nu (q)$ of a quaternion $q$.  Explicitly, it is defined by setting $\nu
(q)=q\bar q$. The norm $\nu (q)$ is a non-negative real number
and $\nu (q)=0$ if and only if $q=0$. By definition, 
$$|q|=\sqrt
{\nu (q)}=\sqrt{q\bar q}=\sqrt{\alpha^2+\beta^2+\gamma^2+\delta^2}.$$ 

It is interesting that every quaternionic number $q $ can be
written as $q =c_1  +\bold jc_2$ where $c_1,c_2\in
\Bbb C$. Moreover, let $M(n, F)$ be the algebra of $n\times n$-matrices over a field
$F$, there has a map $g :\Bbb Q\rightarrow M(2, \Bbb C)$ by
setting
$$
g (q)=\left ( \begin{matrix}
c_1 & -\bar c_2 \\
c_2 & \bar c_1
\end{matrix} \right ).
$$
It is well-known that $g$ is a homomorphism of $\Bbb R$-algebras, and a relation is satisfied $\det (g (q))=\nu (q)$.

Let matrix $Q=(q_{ij})$, $i,j=1,\dots , n$ be a quaternionic
matrix, one can define its corresponding its Hermitian dual matrix to be
$\bar Q^{\prime}=(p_{ij})$ where $p_{ij}=\bar q_{j,i}$ for all $i,j$. Similar to the definition of complex matrix, $Q$ is a
symmetric matrix if $Q= Q^{\prime}$, and $Q$ is a
Hermitian matrix if $Q=\bar Q^{\prime}$, and so on.

\section{Integrals on hyperbolic matrix spaces}
\subsection{Volume of $m\times n$ quaternionic matrix space}
Let  $Q=(q_{ij})$, $i=1,2,\cdots, m$, $j =1,2,\cdots, n$ be a $m\times n$ matrix over a quaternionic field. Explicit expression of the element $q_{ij}$ is
\be
q_{ij}=q^1_{ij}+{\bold i} q^2_{ij}+{\bold j} q^3_{ij}+{\bold k} q^4_{ij},
\ee
where $q^k_{ij}\in \Bbb R, (k=1,2,3,4)$.
The measure of the integral is defined by
$$
\dot{Q}=dQ=\prod_{i=1}^{m}\prod_{j=1}^{n}dq_{ij}.
$$

{\textbf{ Lemma 1.}} Let $f(Q)$ denote any function of $Q$, 
\be\label{lemma1}
\dint_{I-Q\bar{Q}^{\prime}>0}f(Q)\dot{Q}=\dint_{I-Q\bar{Q}^{\prime}>0}\left(\det(I-Q_{m,n-1}\bar{Q}^{\prime}_{m,n-1})\right)^2\dot{Q}_{m,n-1}\dint_{I-w\bar{w}^{\prime}>0}f(Q)\dot{w},
\ee
where $\det$ denotes determinant of the square matrix, and $\dot{w}=\prod_{i=1}^{m}dw_{i}$ relates to the measure of the last column.

\textbf{Proof.} Let us divide $Q$ in the following form
$$
Q=(Q_{m,n-1},q),
$$
where $q$ is the last column of $Q$ and $Q_{{m,n-1}}$ is the matrix consist of the rest elements. Therefore we have
\be
I-Q\bar{Q}^{\prime}=I-Q_{m,n-1}\bar{Q}^{\prime}_{m,n-1}-q\bar{q}^{{\prime}}.
\ee
From the fact that $I-Q\bar{Q}^{\prime}>0$ and $q\bar{q}^{\prime}>0$ we have $I-Q_{m,n-1}\bar{Q}^{\prime}_{m,n-1}>0$. As a consequence, $\exists$ a nonsingular square matrix $\Gamma$ so that 
\be
I-Q_{m,n-1}\bar{Q}^{\prime}_{m,n-1}=\Gamma\bar{\Gamma}^{{\prime}}.
\ee
Under transformation $q=\Gamma w$, we have
\be
\dot{q}=(\det\Gamma\bar{\Gamma}^{\prime})^{2}\dot{w}=(\det(I-Q_{m,n-1}\bar{Q}^{\prime}_{m,n-1}))^{2}\dot{w}.
\ee
Using the relation
\be
I-Q_{m,n-1}\bar{Q}^{\prime}_{m,n-1}-q\bar{q}^{\prime}=\Gamma(I-w\bar{w}^{\prime})\bar{\Gamma}^{\prime},
\ee
one leads to the formula \eqref{lemma1}. The lemma is proved.

{\textbf {Theorem 1.}} Let $Q=Q^{(m,n)}$ is a $m\times n$ quaternionic matrix, $\lambda>-1$ and denote the integral by
$$
J_{m,n}(\lambda)=\dint_{I-Q\bar{Q}^{\prime}>0}\det(I-Q\bar{Q}^{\prime})^{\lambda}\dot{Q},
$$
Then
\be
J_{m,n}(\lambda)=\pi^{2mn}\frac{\prod_{i=1}^{n}\Gamma(\lambda+2j-1)\prod_{k=1}^{m}\Gamma(\lambda+2k-1)}{\prod_{\ell=1}^{n+m}\Gamma(\lambda+2\ell-1)}.
\ee
Particularly, $\lambda=0$ gives the volume of $\Re_{I}$,
\be
Vol(\Re_I)=\pi^{2mn}\frac{\prod_{i=1}^{n}\Gamma(2j-1)\prod_{k=1}^{m}\Gamma(2k-1)}{\prod_{\ell=1}^{n+m}\Gamma(2\ell-1)}.
\ee

\textbf{Proof.} Applying Lemma 1 repeatly, we have
\bea
\nonumber\dint_{I-Q\Bp{Q}>0}f(Q)\dot{Q}&=&\dint_{\Bp{w}_{1}w_{1}<1}(1-\Bp{w}_{1}w_{1})^{2(n-1)}\dot{w}_{1}\dint_{\Bp{w}_{2}w_{2}<1}(1-\Bp{w}_{2}w_{2})^{2(n-2)}\dot{w}_{2}\times\\
&&\times\cdots\times\dint_{\Bp{w}_{n}w_{n}<1}f(Q)\dot{w}_{n}
\eea
Let $f(Q)=(\det(I-Q\Bp{Q}))^{\lambda}$, one has
\bea
\nonumber J_{m,n}(\lambda)&=&\dint_{I-Q\Bp{Q}>0}(\det(I-Q\Bp{Q}))^{\lambda}\dot{Q}\\
&=&\prod_{j=1}^{n}\dint_{\Bp{w}w<1}(1-\Bp{w}_{n-j+1}w_{n-j+1})^{\lambda+2(j-1)}\dot{w}_{n-j+1}.
\eea
Using above equation and the relation
\be\label{int1}
\dint_{x_{1}^{2}+x_{2}^{2}+\cdots+x_{4m}^{2}<1}\left(1-x_{1}^{2}-x_{2}^{2}-\cdots-x_{4m}^{2}\right)^{\mu-1}dx_{1}dx_{2}\cdots dx_{4m}=\pi^{2m}\frac{\Gamma(\mu)}{\Gamma(\mu+2m)},\ \text{for $\mu>0$} 
\ee
one obtains
\bea
\nonumber J_{m,n}(\lambda)&=&\prod_{j=1}^{n}\frac{\pi^{2m} \Gamma(2j+\lambda-1)}{\Gamma(2j+\lambda+2m-1)}\\
&=&\pi^{2mn}\frac{\prod_{i=1}^{n}\Gamma(\lambda+2j-1)\prod_{k=1}^{m}\Gamma(\lambda+2k-1)}{\prod_{\ell=1}^{n+m}\Gamma(\lambda+2\ell-1)}.
\eea
This ends the proof.

{\textbf {Theorem 2.}} Let $Q=Q^{(m,n)}$ be a $m\times n$ quaternionic matrix, $\alpha>2m+2n-2$ and denote the integral by
$$
K_{m,n}(\alpha)=\dint_{Q}\frac{\dot{Q}}{\det(I+Q\bar{Q}^{\prime})^{\alpha}},
$$
Then
\be
K_{m,n}(\alpha)=\pi^{2mn}\prod_{j=0}^{n-1}\frac{\Gamma(\alpha-2j-2m)}{\Gamma(\alpha-2j)}.
\ee

\textbf{Proof.}  Following what we did in proving the Theorem 1, noticing that 
\be
\dint_{Q}\left(1+x_{1}^{2}+x_{2}^{2}+\cdots+x_{4m}^{2}\right)^{-\alpha}dx_{1}dx_{2}\cdots dx_{4m}=\pi^{2m}\frac{\Gamma(\alpha-2m)}{\Gamma(\alpha)},\ \text{for $\alpha>2m$} ,
\ee
we can easily prove the theorem.

\subsection{Hermitian matrix}
\subsubsection{Integral of arctan-like function over quaternionic Hermitian matrix spaces}
As a warmup, let us first consider integral of arctan-like function over quaternionic Hermitian matrix spaces.\\
\textbf{Theorem 3.} If $Q$ is a $n\times n$ Hermitian quaternionic matrix, and $\alpha>2n-\frac32$, then
\bea
\nonumber H_{n}(\alpha)&=&\dint_{Q}\frac{\dot{Q}}{(\det(I+Q^{2}))^{\alpha}}\\
&=& 2^{n(n-1)}\pi^{n\left(n-\frac12\right)}\prod_{j=0}^{n-1}\frac{\Gamma(\alpha-2j-\frac12)}{\Gamma(\alpha-2j)}\prod_{k=0}^{n-2}\frac{\Gamma(2\alpha-2n-2k+1)}{\Gamma(2\alpha-4k-1)}.
\eea

Before proceeding, let us introduce two important lemmas.\\
\textbf{Lemma 2.} Let $u$ and $w$ be $n$-dimensional vectors over a quaternionic field, then \\
(i)
\be
(I+\Bp{u}u)^{-1}\Bp{u}=\frac{\Bp{u}}{1+u\Bp{}u},\ \ \  u(I+\Bp{u}u)^{-1}=\frac{u}{1+u\Bp{}u}
\ee
(ii)
\be
w(I+\Bp{u}u)^{-1}\Bp{w}=w\Bp{w}-\frac{|w\Bp{u}|^2}{1+u\Bp{}u}.
\ee

\textbf{Proof.} (i) Since $u\Bp{u}$ is real number, we therefore have
\be
(u\Bp{u}) \Bp{u}u=\Bp{u}(u\Bp{u}) u,
\ee
which leads to
\be
 \left(1-\frac{1}{1+u\Bp{u}}\right)\Bp{u}u=\Bp{u}u\frac{\Bp{u} u}{1+u\Bp{u}}.
\ee
Direct calculation shows
\be
(I+\Bp{u}u)^{-1}\Bp{u}=\frac{\Bp{u}}{1+u\Bp{}u}.
\ee
Similarly, we have 
\be\label{lemma21}
u(I+\Bp{u}u)^{-1}=\frac{u}{1+u\Bp{}u}.
\ee
(ii) From the identity
\be
w(I+u\Bp{u})=w+(w\Bp{u})u,
\ee
we have
\bea
\nonumber w\Bp{w}&=&w(I+u\Bp{u})(I+u\Bp{u})^{-1}\Bp{w}\\
\nonumber&=&(w+(w\Bp{u})u)(I+u\Bp{u})^{-1}\Bp{w}\\
\nonumber&=& w(I+u\Bp{u})^{-1}\Bp{w}+(w\Bp{u})u(I+u\Bp{u})^{-1}\Bp{w}\\
&=& w(I+u\Bp{u})^{-1}\Bp{w}+\frac{|w\Bp{u}|^2}{1+u\Bp{}u}.
\eea
In the last step, we have used \eqref{lemma21}. Therefore the Lemma is proved.

\textbf{Lemma 3.} Let $a, b, c, \alpha$ are real, and $a>0$, $ac-b^2>0, \alpha>\frac12$, then\\
\be
\int_{-\infty}^{\infty}\frac{dx}{(ax^2+2bx+c)^{\alpha}}=a^{\alpha-1}(ac-b^2)^{\frac12-\alpha}\frac{\sqrt{\pi}\Gamma(\alpha-\frac12)}{\Gamma(\alpha)}.
\ee

\textbf{Proof.} Following Ref. \cite{Hua1}, let 
$$
y=\frac{a}{\sqrt{ac-b^2}}\left(x+\frac{b}{a}\right),
$$
then
\be
\int_{-\infty}^{\infty}\frac{dx}{(ax^2+2bx+c)^{\alpha}}=a^{\alpha-1}(ac-b^2)^{\frac12-\alpha}\frac{\sqrt{\pi}\Gamma(\alpha-\frac12)}{\Gamma(\alpha)}.
\ee
The lemma 3 is proved.

\textbf{Proof of Theorem 3.} Now we can proceed to prove the theorem 3. Dividing
\be
Q=\left(\begin{array}{cc}Q_{1} & \Bp{v} \\v & h\end{array}\right),
\ee
where $h=h_{nn}$ is real, and $Q_{1}$ is $(n-1)\times (n-1)$ quaternionic Hermitian matrix, while $v$ denotes quaternionic vector with $(n-1) $ dimensions. With this division, we have
\be
I+Q^{2}=\left(\begin{array}{cc}I+Q_{1}^{2}+\Bp{v}v  & Q_{1}\Bp{v}+\Bp{v} h \\vQ_{1}+hv &1+h^{2}+v\Bp{v}\end{array}\right),
\ee
Taking the following relation into consider \cite{Hua1},
\be\label{diag}
\left(\begin{array}{cc} I & 0 \\-PA^{-1} & 1 \end{array}\right) \left(\begin{array}{cc} A & \Bp{P} \\ P  & l \end{array}\right) \Bp{\left(\begin{array}{cc} I & 0 \\-PA^{-1} & 1 \end{array}\right) }=\left(\begin{array}{cc} A & 0 \\ 0& l-PA^{-1}\Bp{P} \end{array}\right) ,
\ee
we have
\bea
\nonumber\det (I+Q^2)&=&\left[1+h^2+v\Bp{v}-(vQ_1+hv)(I+Q_1^2+\Bp{v}v)^{-1}(Q_1\Bp{v}+\Bp{v}h)\right]\det(I+Q_1^2+\Bp{v}v)\\
&=& (ah^2+2bh+c)\det(I+Q_1^2+\Bp{v}v),
\eea
where
\bea\label{a1}
a &=& 1-v(I+Q_1^2+\Bp{v}v)^{-1}\Bp{v},\\
2b &=& -vQ_1 (I+Q_1^2+\Bp{v}v)^{-1} \Bp{v}-v(I+Q_1^2+\Bp{v}v)^{-1}Q_1\Bp{v},\\
\label{c1}c&=&1+v\Bp{v}-vQ_1(I+Q_1^2+\Bp{v}v)^{-1}Q_1\Bp{v}.
\eea
 Since $Q_1$ is Hermitian matrix, there exists one unitary matrix $U$ such that
\be
Q_1=U[\lambda_1,\lambda_2,\cdots,\lambda_n]\Bp{U}.
\ee
Therefore another Hermitian matrix can be introduced
\be
T=U[\sqrt{1+\lambda_1^2},\sqrt{1+\lambda_2^2},\cdots, \sqrt{1+\lambda_n^2}]\Bp{U},
\ee
and it satisfies 
\be
TQ_1=Q_1T,\ \ \ I+Q_1^2=T^2.
\ee
After a transfermation $v=uT$,  $a, b, c$ in \eqref{a1}-\eqref{c1} can be rewritten as
\bea\label{a1}
a &=& \frac1{1+u\Bp{u}},\\
2b &=& -\frac{2uQ_1\Bp{u}}{1+u\Bp{u}},\\
\label{c1}c&=&1+u\Bp{u}+\frac{|uQ_1\Bp{u}|}{1+u\Bp{u}},
\eea
where we have used the Lemma 2. It is easy to show $b$ is real and therefore we have the relation
\be
ac-b^2=1.\label{abc}
\ee
By Lemma 3 and Eq. \eqref{abc}, we have the following recursive relation
\bea
\nonumber H_n(\alpha)&=&\dint_{Q}\frac{\dot{Q}}{(\det(I+Q^2))^{\alpha}}\\
\nonumber&=&2^{2(n-1)}\dint_{u,Q_1}(\det(I+Q_1^2))^{2-\alpha}(1+u\Bp{u})^{-\alpha}\dot{Q}_1\dot{u}\int_{-\infty}^{\infty}(ah^2+2bh+c)^{-\alpha}dh\\
&=&4^{n-1}\frac{\pi^{2n-\frac32}\Gamma(2\alpha-2n+1)\Gamma(\alpha-\frac12)}{\Gamma(\alpha)\Gamma(2\alpha-1)}H_{n-1}(\alpha-2)
\eea
Considering 
\be
H_1(\alpha-2n+2)=\int_{-\infty}^{\infty}\frac{dx}{(1+x^2)^{\alpha-2n+2}}=\frac{\sqrt{\pi}\Gamma(\alpha-2n+\frac32)}{\Gamma(\alpha-2n+2)}, \ \ \ \left(\alpha>2n-\frac32\right)
\ee
we obtain
\be
H_n(\alpha)=2^{n(n-1)}\pi^{n\left(n-\frac12\right)}\prod_{j=0}^{n-1}\frac{\Gamma(\alpha-2j-\frac12)}{\Gamma(\alpha-2j)}\prod_{k=0}^{n-2}\frac{\Gamma(2\alpha-2n-4k+1)}{\Gamma(2\alpha-4k-1)}.
\ee
This ends the proof.

\subsubsection{Volume of Hermitian hyperbolic matrix spaces over a quaternion field}
\textbf{Theorem 4.} If $Q$ is a $n\times n$ Hermitian quaternionic matrix, and $\lambda>-1$, then
\bea
\nonumber I_{n}(\lambda)&=&\dint_{I-Q^2>0}(\det(I-Q^2))^{\lambda}\dot{Q}\\
&=& \pi^{n\left(n-\frac12\right)}\prod_{j=0}^{n-1}\frac{\Gamma(\lambda+2j+1)}{\Gamma(\lambda+2j+\frac32)}\prod_{k=0}^{n-2}\frac{\Gamma(2\lambda+4k+2)}{\Gamma(2\lambda+2n+4k)}.
\eea
Particularly, when $\lambda=0$, this gives volume of Hermitian hyperbolic matrix spaces over a quaternionic field
\bea
Vol(\Re_{II})= \pi^{n\left(n-\frac12\right)}\prod_{j=0}^{n-1}\frac{\Gamma(2j+1)}{\Gamma(2j+\frac32)}\prod_{k=0}^{n-2}\frac{\Gamma(4k+2)}{\Gamma(2n+4k)}.
\eea

\textbf{Proof.} Following what we did in Theorem 3 and noticing  that
\be
\int\limits_{ax^2+2bx+c>0}(ax^2+2bx+c)^{\lambda}dx=(-a)^{-\lambda-1}\frac{\sqrt{\pi}\Gamma(\lambda+1)}{\Gamma(\lambda+\frac32)}, \ \ \ \  (\lambda>-1),
\ee
and a integral followed by \eqref{int1} 
\be
\dint_{1-u\Bp{u}>0}(1-u\Bp{u})^{2\lambda+1}\dot{u}=\frac{\pi^{2(n-1)}\Gamma(2\lambda+2)}{\Gamma(2\lambda+2n)},
\ee
we can obtain a recursive relation
$$
I_{n}(\lambda)=\frac{\pi^{2n-\frac32}\Gamma(\lambda+1)\Gamma(2\lambda+2)}{\Gamma(\lambda+\frac32)\Gamma(2\lambda+2n)}I_{n-1}(\lambda+2).
$$
Applying this formula repeatly, we get the final result. This ends the proof.

\subsection{Volume of symmetric square matrix spaces over a quaternionic field}
\textbf{Lemma 4.} Let $a, c \in \Bbb R$, $b\in \Bbb Q$, and $a<0$, $|b|^2-ac>0$, $\lambda>-1$, then
\be
\dint_{c+b\bar{q}+q\bar{b}+qa\bar{q}>0}\left(c+b\bar{q}+q\bar{b}+qa\bar{q}\right)^{\lambda}\dot{q}=\frac1{a^2}\left(\frac{|b|^2-ac}{|a|}\right)^{\lambda+2}\frac{\pi^2}{(\lambda+1)(\lambda+2)}.
\ee
\textbf{Proof.} Let
$$
w=\left(q+\frac{b}{a}\right)\sqrt{\frac{a^2}{|b|^2-ac}},
$$
which implies
$$
\dot{w}=\left(\frac{a^2}{|b|^2-ac}\right)^2\dot{q},
$$
and 
$$
c+b\bar{q}+q\bar{b}+qa\bar{q}=\frac{|b|^2-ac}{|a|}(1-w\bar{w}).
$$
As a consequence, we obtain
\bea
\nonumber&&\dint_{c+b\bar{q}+q\bar{b}+qa\bar{q}>0}\left(c+b\bar{q}+q\bar{b}+qa\bar{q}\right)^{\lambda}\dot{q}\\
\nonumber&&=\frac1{a^2}\left(\frac{|b|^2-ac}{|a|}\right)^{\lambda+2}\iint\limits_{1-w\bar{w}>0}(1-w\bar{w})^{\lambda}\dot{w}\\
&&=\frac1{a^2}\left(\frac{|b|^2-ac}{|a|}\right)^{\lambda+2}\frac{\pi^2}{(\lambda+1)(\lambda+2)}.
\eea
The Lemma is proved.\\
\textbf{Theorem 5.}  Let $Q=Q^{(n\times n)}=Q^{\prime}$ be a symmetric square matrix over the quaternionic field, and 
$$
J_{n}(\lambda)=\dint_{I-Q\bar{Q}>0}(\det (I-Q\bar{Q}))^{\lambda}\dot{Q}
$$
be the volume, then for $\lambda>-1$ ,
\be
J_n(\lambda)=\frac{\pi^{n(n+1)}}{(\lambda+1)\cdots (\lambda+2n)}\frac{\Gamma(2\lambda+5)\cdots\Gamma(2\lambda+4n-3)}{\Gamma(2\lambda+2n+3)\cdots\Gamma(2\lambda+4n-1)}.
\ee
A special case $\lambda=0$ gives the volume of the symmetric square matrix on the hyperbolic matrix space  over a quaternionic field, and is given by
\be
Vol(S)=\frac{\pi^{n(n+1)}}{(2n)!}\frac{4!8!\cdots (4n-4)!}{(2n+2)!(2n+4)!\cdots (4n-2)!}.
\ee
\textbf{Proof.} Let
\be
Q=\left(\begin{array}{cc}Q_{1} &v' \\v & q\end{array}\right),
\ee
where $Q_1$ is $(n-1)\times (n-1)$ a symmetric square matrix over the field, $v$ is an $(n-1)$ vector and $q$ is a quaternion. After doing so, we find
\be
I-Q\bar{Q}=\left(\begin{array}{cc}I-Q_{1}\bar{Q}_1-v' \bar{v} & -(Q_1\Bp{v}+v' \bar{q}) \\-(v\bar{Q_1}+q\bar{v} & 1-v\Bp{v}-q\bar{q}\end{array}\right),
\ee
Eq. (\ref{diag}) implies that 
\be
\det(I-Q\bar{Q})=(c+b\bar{q}+q\bar{b}+qa\bar{q})\det(I-Q_1\bar{Q}_1-v' \bar{v}),
\ee
and $c+b\bar{q}+q\bar{b}+qa\bar{q}>0, \det(I-Q_1\bar{Q}_1-v' \bar{v})>0$ 
where
\bea
a&=&-1-\bar{v}(I-Q_1\bar{Q}_1-v' \bar{v})^{-1}v',\\
b&=&-v\bar{Q}_1(I-Q_1\bar{Q}_1-v' \bar{v})^{-1}v',\\
c&=&1-v\Bp{v}-v\bar{Q}_1(I-Q_1\bar{Q}_1-v' \bar{v})^{-1}Q_1\Bp{v}.
\eea
Moreover, since $I-Q_1\bar{Q}_1$ is positive definite, there exists a nonsingular square matrix $\Gamma$ such that $I-Q_1\bar{Q}_1=\Gamma\Bp{\Gamma}$.
After introducing a transformation $v=u\Gamma'$, we obtain the following results
\bea\label{a2}
a&=&-\frac1{1-\bar{u}u'} (<0),\\
b&=&-\frac{u\Gamma'\bar{Q}_1\left(\Bp{\Gamma}\right)^{-1}u'}{1-\bar{u}u'},\\
\label{c2}c&=&1-u\Gamma' \bar{\Gamma}\Bp{u}-u\Gamma' \bar{Q}_1\left(\Bp{\Gamma}\right)^{-1}\Gamma^{-1}\bar{Q}_1\bar{\Gamma}\Bp{u}-\frac{|u\Gamma'\bar{Q}_1\left(\Bp{\Gamma}\right)^{-1}u'|^2}{1-\bar{u}u'},
\eea
which leads to $|b|^2-ac=1$.

In deriving \eqref{a2}-\eqref{c2}, we have used the following relations which can be proved the same way as given in Lemma 2
\bea
(I-u' \bar{u})^{-1}u'=\frac{u'}{1-\bar{u}u'},\\
w(I-u' \bar{u})^{-1}\Bp{w}=w\Bp{w}+\frac{|wu'|^2}{1-\bar{u}u'}.
\eea
Using these equations and Lemma 4, we have
\bea
\nonumber J_n(\lambda)&=&\dint_{I-Q_1\bar{Q}_1-v' \bar{v}>0}(\det(I-Q_1\bar{Q}_1-v' \bar{v}))^{\lambda}\dot{Q}_1\dot{v}\dint_{c+b\bar{q}+q\bar{b}+qa\bar{q}>0}(c+b\bar{q}+q\bar{b}+qa\bar{q})^{\lambda}\dot{q},\\
\nonumber&=&\frac{\pi^2}{(\lambda+1)(\lambda+2)}\dint_{I-Q_1\bar{Q}_1>0}(\det(I-Q_1\bar{Q}_1))^{\lambda+2}\dot{Q}_1\dint_{\bar{u}u'<1}(1-\bar{u}u')^{2\lambda+4}\dot{u},\\
&=& \frac{\pi^{2n}}{(\lambda+1)(\lambda+2)}\frac{\Gamma(2\lambda+5)}{\Gamma(2\lambda+2n+3)}J_{n-1}(\lambda+2).
\eea
Repeating this calculation and noticing that for $n=1$, there has 
$$
J_1(\lambda)=\frac{\pi^2}{(\lambda+1)(\lambda+2)},
$$
we therefore get the final result. This ends the proof.

\subsection{Volume of anti-Hermitian hyperbolic matrix spaces over a quaternion field}
\textbf{Lemma 5.} 
Let $Q=Q^{\prime}=Q_0+\bold i Q_1+\bold j Q_2+\bold k Q_3$ be a $n\times n$($n>1$) symmetric square matrix over the quaternionic field with $Q_3=-(Q_1+Q_2)$, then for $\lambda>-1$ ,
\be
\dint_{Q}(\det (I-Q\bar{Q}))^{\lambda}\dot{Q}= \left(\frac{\pi^3}{3}\right)^{\frac{n(n+1)}{4}}\frac{\Gamma(\lambda+1)}{\Gamma(\lambda+\frac{3n}2+1)}\cdot\prod_{j=1}^{n-1}\frac{\Gamma(2\lambda+3j+1)}{\Gamma(2\lambda+\frac{3(n+j)}{2}+1)}.
\ee
\textbf{Proof. } Following the same procedures  as what we did in theorem 5, and being aware that the Lemma 4 in the present case should be replaced by 
\be
\dint_{c+b\bar{q}+q\bar{b}+qa\bar{q}>0}\left(c+b\bar{q}+q\bar{b}+qa\bar{q}\right)^{\lambda}\dot{q}=\left(\frac{\pi}{|a|}\right)^{\frac32}\left(\frac{|b|^2-ac}{|a|}\right)^{\lambda+\frac32}\frac{\Gamma(\lambda+1)}{\sqrt{3}\Gamma(\lambda+\frac52)},
\ee
where  $a, c \in \Bbb R$, $b\in \Bbb Q$, and $a<0$, $|b|^2-ac>0$, $\lambda>-1$, one can easily prove it.

\textbf{Theorem 6.} If $H$ is an $n\times n$ anti-Hermitian quaternionic matrix, and $\lambda>-1$, then
\bea
\nonumber K_{n}(\lambda)&=&\dint_{I-H\Bp{H}>0}(\det(I-H\Bp{H}))^{\lambda}\dot{H}=\dint_{I+H^2>0}(\det(I+H^2))^{\lambda}\dot{H}\\
&=& \left(\frac{\pi^3}{3}\right)^{\frac{n(n+1)}{4}}\frac{\Gamma(\lambda+1)}{\Gamma(\lambda+\frac{3n}2+1)}\cdot\prod_{j=1}^{n-1}\frac{\Gamma(2\lambda+3j+1)}{\Gamma(2\lambda+\frac{3(n+j)}{2}+1)}.\eea
Particularly, when $\lambda=0$, this gives volume of $\Re_{III}$ over a quaternionic field
\bea
Vol(\Re_{III})=  \left(\frac{\pi^3}{3}\right)^{\frac{n(n+1)}{4}}\frac{1}{\Gamma(\frac{3n}2+1)}\cdot\prod_{j=1}^{n-1}\frac{\Gamma(3j+1)}{\Gamma(\frac{3(n+j)}{2}+1)}.\eea

\textbf{Proof.}  Let us define an $n\times n$ matrix $Q=Q_0+\bold i Q_1+\bold j Q_2+\bold k Q_3$ ($Q_0,Q_1,Q_2,Q_3$ are real matrices) in the following way,
\be\label{antiHermite1}
H=\frac{Q}{\sqrt{3}}(\bold i+\bold j+\bold k) .
\ee
Since $H$ is an anti-Hermitian matrix, it is easy to check that
\be\label{antiHermite2}
Q(\bold i+\bold j+\bold k) =(\bold i+\bold j+\bold k)\Bp{Q},
\ee
which implies that
\be\label{antiHermite3}
Q'=Q, \ \text{and},\ \ Q_3=-(Q_1+Q_2).
\ee

As a consequence, 
\bea
\nonumber K_{n}(\lambda)=\dint_{I+H^2>0}(\det(I+H^2))^{\lambda}\dot{H}=\dint_{Q}(\det(I-Q\bar{Q}))^{\lambda}\dot{Q}.
\eea
From Lemma 5 we get
\be
K_n(\lambda)=\dint_{Q}(\det (I-Q\bar{Q}))^{\lambda}\dot{Q}= \left(\frac{\pi^3}{3}\right)^{\frac{n(n+1)}{4}}\frac{\Gamma(\lambda+1)}{\Gamma(\lambda+\frac{3n}2+1)}\cdot\prod_{j=1}^{n-1}\frac{\Gamma(2\lambda+3j+1)}{\Gamma(2\lambda+\frac{3(n+j)}{2}+1)}.
\ee
This ends the proof.

\subsection{Volume of $\Re_{IV}$ over a quaternionic field}

Let $q$ be $n$-dimensional quaternionic vector $(q_1, q_2,\cdots, q_n)$. The fourth classical domain $\Re_{IV}$ is a set of $q$ which satisfies 
$$
|qq'|^2+1-2\bar{q}q'>0\ \ \ \text{and} \ \ |qq'|<1,
$$
or equivalently 
\be
1-\bar{q}q'>\sqrt{(\bar{q}q')^2-|qq'|^2}.
\ee

\textbf{Lemma 6.} If $a,b\in \Bbb R$, $0<a<1, b>-1$, then
\be
\mathop{\iint\limits_{x\geqslant0,y\geqslant 0}}_{x^2+y^2\leqslant a^2} (a^2-x^2-y^2)^bx^{2b+1}y^{2b+1}dxdy=\frac{a^{6b+4}\sqrt{\pi}}{2^n}\frac{\Gamma(b+1)^2\Gamma(2b+2)}{\Gamma(b+\frac{3}{2})\Gamma(3b+3)}.
\ee
\textbf{Proof.}  Defining $\hat{x}=\frac{x}{a}, \hat{y}=\frac{y}{a}$,
\bea
\nonumber\mathop{\iint\limits_{x\geqslant0,y\geqslant 0}}_{x^2+y^2\leqslant a^2} (a^2-x^2-y^2)^bx^{2b+1}y^{2b+1}dxdy&=& a^{6b+4}\mathop{\iint\limits_{\hat{x}\geqslant0,\hat{y}\geqslant 0}}_{\hat{x}^2+\hat{y}^2\leqslant 1}(1-\hat{x}^2-\hat{y}^2)^b\hat{x}^{2b+1}\hat{y}^{2b+1}d\hat{x}d\hat{y}\\
\nonumber&=& \frac{a^{6b+4}}{4} \int_{0}^1dX\int_{0}^{1-X} \Big((1-X-Y)XY\Big)^bdY\\
&=& \frac{a^{6b+4}\sqrt{\pi}}{2^n}\frac{\Gamma(b+1)^2\Gamma(2b+2)}{\Gamma(b+\frac{3}{2})\Gamma(3b+3)}.
\eea
where we have introduced $X=\hat{x}^2, Y=\hat{y}^2$. The Lemma is proved.

\textbf{Theorem 7.} If $\alpha>-1, \beta>-(n+\alpha)$, then
\bea
\nonumber L_n(\alpha,\beta)&=& \dint_{\Re_{IV}}\left(1-\bar{q}q'-\sqrt{(\bar{q}q')^2-|qq'|^2} \right)^{\alpha}\left(1-\bar{q}q'+\sqrt{(\bar{q}q')^2-|qq'|^2} \right)^{\beta}\dot{q}\\
\nonumber&=& \begin{cases}
\frac{\pi^2}{(\alpha+\beta+1)(\alpha+\beta+2)},  &\text{for $n=1$;}\\
\frac{2^{5-4n}\pi^{2n+1}\Gamma(n)\Gamma(\alpha+1)\Gamma(1+n+\alpha+\beta)}{\Gamma(\frac n2)\Gamma(\frac32(n-1))\Gamma(\alpha+n+1)\Gamma(2n+\alpha+\beta+1)\Gamma(\frac52-n)}\times\\
\times\sum_{i=1}^{n-1}{2n-3\choose 2i-1}{}_3F_2(1-i,n,1+n+\alpha+\beta; \alpha+n+1,\frac52-n;-1), &\text{for $n\geqslant2$}.
\end{cases}
\eea
In particular, when $\alpha=\beta=0$, it gives the volume of $\Re_{IV}$
\bea
\nonumber Vol(\Re_{IV})= \begin{cases}
\frac{\pi^2}{2},  &\text{for $n=1$;}\\
\frac{2^{5-4n}\pi^{2n+1}\Gamma(n)}{\Gamma(\frac n2)\Gamma(\frac32(n-1))\Gamma(2n+1)\Gamma(\frac52-n)}\times\\
\times\sum_{i=1}^{n-1}{2n-3\choose 2i-1}{}_3F_2(1-i,n,1+n; n+1,\frac52-n;-1), &\text{for $n\geqslant2$}.
\end{cases}
\eea

\textbf{Proof.} (i)  $n=1$case. Let $q=x+\bold i y+\bold j z+\bold k w$, ($x,y,z,w\in\Bbb R$).
We have
$$
\bar{q}q'=|qq'|=x^2+y^2+z^2+w^2,
$$
which leads to
\be
L_1(\alpha,\beta)=\dint_{x^2+y^2+z^2+w^2<1}(1-(x^2+y^2+z^2+w^2))^{\alpha+\beta}dxdydzdw=\frac{\pi^2}{(\alpha+\beta+1)(\alpha+\beta+2)}.
\ee
This case is proved.\\
(ii)  $n\geqslant 2$ case. Let $q=x+\bold i y+\bold j z+\bold k w$, but now $x, y, z, w$ are $n$-dimensional real vectors. We therefore have
\bea
1-qq' &=& 1-(xx'+yy'+zz'+ww'),\\
(\bar{q}q')^2-|qq'|^2&=& 4\Big(xx'(yy'+zz'+ww')-\left((xy')^2+(xz')^2+(xw')^2\right)\Big)
\eea
For any fixed $x$, there exists an orthogonal matrix $R$ with unit determinant such that
$$
xR=(\sqrt{xx'},0,0,\cdots,0).
$$
Meanwhile, we also have 
$$
yR=(\xi,\tilde{y}),\ \ \  zR=(\zeta,\tilde{z}),\ \ \  wR=(\epsilon,\tilde{w}),\ \ \ 
$$
where $\xi,\zeta,\epsilon$ are real numbers, while $ \tilde{y}, \tilde{z}, \tilde{w}$ are $(n-1)$-dim real vectors. After introducing the following transformations
$$(x,\tilde{y},\tilde{z},\tilde{w})=\sqrt{1-\xi^2-\zeta^2-\epsilon^2}   (u,v,r,s),$$
we obtain
\bea
\label{Ln} L_n(\alpha,\beta)=2^{4n-3} \frac{\pi^{\frac32}\Gamma(2n+\alpha+\beta-\frac12)}{\Gamma(2n+\alpha+\beta+1)} P,
\eea
where
\bea
\nonumber P&=&\mathop{\dint_{1-uu'-vv'-rr'-ss'>2\sqrt{uu'(vv'+rr'+ss')}}}_{u_i,v_i,r_i,s_i\geqslant0}\Big(1-uu'-vv'-rr'-ss'-2\sqrt{uu'(vv'+rr'+ss')}\Big)^{\alpha}\times\\
&&\times \Big(1-uu'-vv'-rr'-ss'+2\sqrt{uu'(vv'+rr'+ss')}\Big)^{\beta}\dot{u}\dot{v}\dot{r}\dot{s}.
\eea
Defining $\rho^2=uu',\sigma^2=vv',\kappa^2=rr',\eta^2=ss'$, then we have
\bea
\nonumber \dint_{u_i\geqslant0}\dot{u}&=&\dint_{u_i\geqslant0}du_1du_2\cdots du_n\\
\nonumber&=&\int\rho d\rho \mathop{ \dint_{u_2^2+\cdots+u_n^2\leqslant \rho^2}}_{u_i\geqslant0} \frac{du_2\cdots du_n}{\sqrt{\rho^2-u_2^2-\cdots-u_n^2}}\\
&=&\frac{\pi^{\frac n2}}{2^{n-1}\Gamma(\frac n2)}\int\rho^{n-1}d\rho.
\eea
Similarly we  have (note that $v, r, s$ are $(n-1)$-dimensional vectors)
\bea
\dint_{v_i\geqslant0}\dot{v}&=& \frac{\pi^{\frac {n-1}2}}{2^{n-2}\Gamma(\frac {n-1}2)}\int\sigma^{n-2}d\sigma,\\
\dint_{r_i\geqslant0}\dot{r}&=& \frac{\pi^{\frac {n-1}2}}{2^{n-2}\Gamma(\frac {n-1}2)}\int\kappa^{n-2}d\kappa,\\
\dint_{s_i\geqslant0}\dot{s}&=&  \frac{\pi^{\frac {n-1}2}}{2^{n-2}\Gamma(\frac {n-1}2)}\int\eta^{n-2}dd\eta.
\eea
As a consequence $P$ becomes
\bea
\nonumber P&=&\frac{\pi^{2n-\frac32}}{2^{4n-7}\Gamma(\frac n2)\Gamma(\frac{n-1}{2})^3}\times\\
&&\nonumber\times\mathop{ \mathop{\dint_{\rho+\mu<1}}_{\rho,\mu,\kappa,\eta\geqslant0}}_{\kappa^2+\eta^2\leqslant\mu^2} \Big(1-(\rho+\mu)^2\Big)^{\alpha}\Big(1-(\rho-\mu)^2\Big)^{\beta}\rho^{n-1}\mu(\mu^2-\kappa^2-\eta^2)^{\frac{n-3}{2}}\kappa^{n-2}\eta^{n-2}d\rho d\mu d\kappa d\eta,\\
\eea
where $\mu^2=\sigma^2+\kappa^2+\eta^2$.
By Lemma 6 we obtain
\bea
\nonumber P&=&\frac{\pi^{2n-1}\Gamma(n-1)}{2^{5n-7}\Gamma(\frac n2)^2\Gamma(\frac{n-1}{2})\Gamma(\frac32(n-1))}\mathop{\mathop{\iint}_{\rho+\mu<1}}_{\rho\geqslant0,\mu\geqslant0}\Big(1-(\rho+\mu)^2\Big)^{\alpha}\Big(1-(\rho-\mu)^2\Big)^{\beta}\rho^{n-1}\mu^{3n-4}d\rho d\mu\\
\nonumber&=&\frac{\pi^{2n-1}\Gamma(n-1)}{2^{5n-7}\Gamma(\frac n2)^2\Gamma(\frac{n-1}{2})\Gamma(\frac32(n-1))}\times\\
&&\times\mathop{\mathop{\iint}_{\rho+\mu<1}}_{0\leqslant\rho\leqslant\mu}\Big(1-(\rho+\mu)^2\Big)^{\alpha}\Big(1-(\rho-\mu)^2\Big)^{\beta}\rho^{n-1}\mu^{n-1}\left(\mu^{2n-3}+\rho^{2n-3}\right)d\rho d\mu
\eea
To proceed, we define $\tau=\mu-\rho, \nu=\mu+\rho$, then
\bea
\nonumber P&=&\frac{\pi^{2n-1}\Gamma(n-1)}{2^{9n-11}\Gamma(\frac n2)^2\Gamma(\frac{n-1}{2})\Gamma(\frac32(n-1))}\times\\
\nonumber&&\times\int_0^1(1-\tau^2)^{\beta}d\tau\int_{\tau}^1(1-\nu^2)^{\alpha}(\nu^2-\tau^2)^{n-1}\left[(\nu+\tau)^{2n-3}+(\nu-\tau)^{2n-3}\right]d\nu\\
\nonumber &=& \frac{\pi^{2n-1}\Gamma(n-1)}{2^{9n-11}\Gamma(\frac n2)^2\Gamma(\frac{n-1}{2})\Gamma(\frac32(n-1))}\times\\
\nonumber&&\times\sum_{i=1}^{n-1}{2n-3 \choose 2i-1}\int_0^1(1-\tau^2)^{\beta}(\tau^2)^{n-i-1}d\tau\int_{\tau}^1(1-\nu^2)^{\alpha}(\nu^2-\tau^2)^{n-1}2\nu(\nu^2)^{i-1}d\nu\\
\nonumber &=& \frac{\pi^{2n-1}\Gamma(n)\Gamma(n-1)\Gamma(\alpha+1)}{2^{9n-11}\Gamma(\frac n2)^2\Gamma(\frac{n-1}{2})\Gamma(\frac32(n-1))\Gamma(\alpha+n+1)}\times\\
\label{P}\nonumber&&\times\sum_{i=1}^{n-1}{2n-3\choose 2i-1}\int_0^1 (\tau^2)^{n-2}(1-\tau^2)^{\alpha+\beta+n}{}_2F_1(1-i,n;\alpha+n+1;\frac{\tau^2-1}{\tau^2})d\tau.\\
\eea
After expanding the hypergeometric function one gets
\bea
\nonumber&&\int_0^1 (\tau^2)^{n-2}(1-\tau^2)^{\alpha+\beta+n}{}_2F_1(1-i,n;\alpha+n+1;\frac{\tau^2-1}{\tau^2})d\tau\\
\nonumber&&= \frac{\Gamma(\alpha+n+1)}{\Gamma(n)\Gamma(1-i)}\sum_{k=0}^{\infty}\frac{(-1)^k\Gamma(k+1-i)\Gamma(k+n)}{\Gamma(k+\alpha+n+1)k!}\int_0^1(\tau^2)^{n-2-k}(1-\tau^2)^{k+n+\alpha+\beta}d\tau\\
\nonumber&&=\frac{\Gamma(\alpha+n+1)}{\Gamma(n)\Gamma(1-i)}\sum_{k=0}^{\infty}\frac{(-1)^k\Gamma(k+1-i)\Gamma(k+n)}{\Gamma(k+\alpha+n+1)k!}\times\frac{\Gamma(k+1+n+\alpha+\beta)\Gamma(n-k-\frac32)}{2\Gamma(2n+\alpha+\beta-\frac12)}\\
\nonumber&&= \frac{\pi\Gamma(\alpha+n+1)}{2\Gamma(n)\Gamma(1-i)\Gamma(2n+\alpha+\beta-\frac12)}\sum_{k=0}^{\infty}\frac{(-1)^k\Gamma(k+1-i)\Gamma(k+n)\Gamma(k+1+n+\alpha+\beta)}{\Gamma(k+\alpha+n+1)\Gamma(k+\frac52-n)k!}\\
&&= \frac{\pi\Gamma(1+n+\alpha+\beta)}{2\Gamma(n+\alpha+\beta-\frac12)\Gamma(\frac52-n)}\times{}_3F_2(1-i,n,1+n+\alpha+\beta; \alpha+n+1,\frac52-n;-1).
\eea
Substituting this integral into \eqref{P}, we obtain
\bea
\nonumber P&=&\frac{\pi^{2n-\frac12}\Gamma(n)\Gamma(\alpha+1)\Gamma(1+n+\alpha+\beta)}{2^{8(n-1)}\Gamma(\frac n2)\Gamma(\frac32(n-1))\Gamma(\alpha+n+1)\Gamma(n+\alpha+\beta-\frac12)\Gamma(\frac52-n)}\times\\
&&\times\sum_{i=1}^{n-1}{2n-3\choose 2i-1}{}_3F_2(1-i,n,1+n+\alpha+\beta; \alpha+n+1,\frac52-n;-1)
\eea
After inserting it into \eqref{Ln}, we have
\bea
\nonumber L_n(\alpha,\beta)&=&2^{5-4n}\frac{\pi^{2n+1}\Gamma(n)\Gamma(\alpha+1)\Gamma(1+n+\alpha+\beta)}{\Gamma(\frac n2)\Gamma(\frac32(n-1))\Gamma(\alpha+n+1)\Gamma(2n+\alpha+\beta+1)\Gamma(\frac52-n)}\times\\
&&\times\sum_{i=1}^{n-1}{2n-3\choose 2i-1}{}_3F_2(1-i,n,1+n+\alpha+\beta; \alpha+n+1,\frac52-n;-1)
\eea
This ends the proof.

 \section*{\bf Acknowledgements}
This work is supported in part by the National Natural Science Foundation of China under Grants Nos. 11005165 and 11465012 (F. W. S.), and 11373020 (Y. G. S.);  the Natural Science Foundation of Jiangxi Province (under Grant No. 20142BAB202007) and the 555 talent project of Jiangxi Province (F. W. S.).

\end{document}